\documentstyle[12pt,epsfig]{article}
\newcommand {\be}{\begin{equation}}
\newcommand {\ee}{\end{equation}}
\newcommand {\ba}{\begin{eqnarray}}
\newcommand {\ea}{\end{eqnarray}}

\makeatletter \@addtoreset{equation}{section}

\makeatother

\begin{document}
\begin{flushright}
 \today
\end{flushright}

\begin{center}
{\Large \bf Leptonic CP violation: zero,  maximal or between the two extremes}\\
\vskip 0.5cm

{Yasaman Farzan$^{a}$~\footnote{yasaman@theory.ipm.ac.ir} and
Alexei Yu. Smirnov$^{b,c}$~\footnote{smirnov@ictp.it} }
\\

\vskip 0.2cm

{\it $^{a}$ Institute for Studies in Theoretical Physics and
Mathematics
(IPM), P.O. Box 19395-5531, Tehran, Iran\\
$^b$ International Centre for Theoretical Physics, Strada Costiera
11,
34014 Trieste, Italy, \\
$^c$ Institute for Nuclear Research, Russian Academy of Sciences, Moscow,
Russia\\
 }

\end{center}

\begin{abstract}

Discovery of the CP-violation in the lepton sector is one of the
challenges of the particle physics. We search for possible
principles, symmetries and phenomenological relations that can
lead to particular values of the CP-violating Dirac phase,
$\delta$. In this connection we discuss two extreme cases: the
zero phase, $\delta = 0$, and the maximal CP-violation, $\delta =
\pm \pi/2$,  and relate them to the peculiar pattern of the
neutrino mixing. The maximal CP-violation can be related to the
$\nu_\mu - \nu_\tau$ reflection symmetry. We study various aspects
of this symmetry and introduce a generalized reflection symmetry
that can lead to an arbitrary phase that depends on the parameter
of the symmetry transformation. The generalized reflection
symmetry predicts a simple relation between the Dirac and Majorana
phases. We also consider the possibility of certain relations
between the CP-violating phases in the quark and lepton sectors.

\end{abstract}

\section{Introduction}

Observation of the effects of the CP violation  was one of the
fundamental discoveries in physics~\cite{cp} in the past century.
The violation of the CP-symmetry is established in the quark
sector and it is natural to expect that CP violation occurs in the
lepton sector, too. Furthermore,  in the lepton sector one  may
even find additional sources of  CP-violation; e.g., the  Majorana
phases of neutrino masses, the right handed (RH) neutrino mass
matrix in the context of seesaw mechanism, or mixing with new
neutrino states.

What can we learn from measurements of the CP-violating phases?
What is the underlying physics? These questions are imperative
especially in view of development of the challenging and rather
expensive experimental programs to measure the CP-violation  in
the neutrino oscillations~\cite{factory}. What would the  possible
implications of the future measurements of the phase be?

We still have no  theory of CP-violation in the quark sector that
would explain the observed value of the phase $\delta_{CKM}$. What
can we say  about  the CP-violation in the lepton sector where the
information about masses and mixings is not so complete as in the
quark sector? Can the situation be simpler here?


In view of this incomplete knowledge,
we  can take the following routes to approach the questions raised
above.

1) Some extreme situations can be studied; e.g., the possibility
of zero Dirac phase, $\delta = 0$, or maximal phase, $\delta =
\pi/2$.

2) We can also try to relate the CP-violating phases in the quark
and lepton sectors. In this line, one may ask if  the phases can
be equal or complementary; or if there is another  simple relation
between the phases.

There are few specific  models of neutrino mass and mixing that
predict  the value of the CP-violating phase. For example, the Zee
model (whose minimal version is excluded by the recent data)
predicts a neutrino mass matrix invariant under CP. In the models
with $A_4 $ symmetry the maximal  value for the Dirac phase,
$\delta = \pi/2$, appears \cite{A4CP}. Some structures of the mass
matrices lead to maximal mixing~\cite{yasue}. Classification of
matrices with a certain number of texture zeros in the flavor
basis and the corresponding CP-violation effects have been
considered in~\cite{texturezero}. However, till now no systematic
study of the CP-violation exists. In this paper, we search for
principles, symmetries as well as  phenomenological and empirical
relations that  lead to particular values of the CP-violating
phase. In sec. 2, we study some general properties of the neutrino
mass matrix and formulate criteria for the CP-violation. In sec.
3, we search for symmetries which predict $\delta = 0$ and discuss
the related phenomenological consequences. In sec. 4, we consider
the case of maximal CP-violation. In sec. 5, we formulate
conditions that lead to certain values of the phase which differ
from zero and $\pi/2$. The possibility of relations between the
quark and lepton CP-violating phases is studied in sec. 6. In sec.
7, we summarize our results.

\section{Rephasing invariants and criteria for conservation of CP}

We assume that there are only three light Majorana neutrinos and
parameterize the neutrino mass matrix in the flavor basis $\nu_f
\equiv (\nu_e, \nu_{\mu}, \nu_{\tau})$  as \ba m_\nu=\left[
\matrix{ m_{ee} & m_{e\mu} & m_{e\tau} \cr ... & m_{\mu\mu} &
m_{\mu \tau} \cr ... & ... & m_{\tau\tau} }\right]. \label{massm}
\ea
We will use the standard parametrization of the mixing matrix
which diagonalizes $m_{\nu}$ \be U_{PMNS} = U_{23}(\theta_{23})
\Gamma_{\delta}U_{13}(\theta_{13}) \Gamma_{-\delta}
U_{12}(\theta_{12}) \Gamma_M. \ee Here $U_{ij}(\theta_{ij})$ is
the matrix of rotation by an angle $\theta_{ij}$ in the
$ij$-plane; \be \Gamma_{\delta} \equiv {\rm diag} (1, 1,
e^{i\delta}),~~~ \Gamma_M \equiv {\rm diag} (1, e^{i\phi_2/2},
e^{i\phi_3/2}), \ee where $\phi_2$, $\phi_3$ are the Majorana
phases defined in such a way that mass eigenvalues are made real
and positive. These phases could be included in the definition of
the eigenvalues $m_i$, and since the oscillation probabilities are
determined by $|m_i|^2$, only the phase $\delta$ affects
oscillations.

Let us introduce the following matrix \be h \equiv m_\nu \cdot
m_\nu^\dagger. \label{hmatr} \ee The CP-violation in neutrino
oscillations is then determined by \cite{Jarlskog} \be {\cal J}=
{\rm Im}[h_{e\mu} h_{\mu\tau} h_{\tau e} ] \label{i-expr} \ee
which is related to the Jarlskog invariant, $J_{CP}$, as \be {\cal
J} = \Delta m_{12}^2 \Delta m_{32}^2 \Delta m_{13}^2 J_{CP}.
\label{calj} \ee Here \be \label{JCP} J_{CP} =
s_{12}c_{12}s_{23}c_{23} s_{13}c_{13}^2\sin\delta \ee and  $\Delta
m_{ij}^2 \equiv m_i^2 - m_j^2$, $s_{12} \equiv \sin \theta_{12}$,
$c_{12} \equiv \cos \theta_{12}$, {\it etc}. In terms of $m_\nu$
the elements $h_{\alpha \beta}$  are given by \ba h_{e\mu} &=&
m_{ee} m_{e\mu}^* + m_{e\mu} m_{\mu\mu}^* + m_{e\tau}m_{\tau
\mu}^* \cr h_{\mu \tau} &=& m_{\mu e} m_{e\tau}^* + m_{\mu \mu}
m_{\mu\tau}^* + m_{\mu\tau}m_{\tau \tau}^* \cr h_{\tau e} &=&
m_{\tau e} m_{ee}^* + m_{\tau \mu} m_{\mu e}^* + m_{\tau
\tau}m_{\tau e}^* . \ea Explicit expressions for $h_{\alpha
\beta}$ in terms of oscillation parameters are given in the
appendix. Notice that $h$ does not depend on the Majorana phases
and therefore provides a test of the Dirac phase without any
ambiguity from the Majorana phases.

In order to investigate symmetries and relations that determine
the CP-violation in general (either through the Majorana phases or
through the Dirac phase) we should consider the matrix $m_\nu$
rather than $h$. However, the arguments of the elements of $m_\nu$
change with rephasing the fields and therefore are not physical.
To perform such an analysis, we should use  rephasing invariant
quantities~\cite{shrock}. Under rephasing of neutrino states,
$(\nu_e, \nu_{\mu}, \nu_{\tau}) \rightarrow (e^{i\alpha_e} \nu_e,
e^{i\alpha_\mu} \nu_{\mu}, e^{i\alpha_\tau}\nu_{\tau})$, the
matrix changes as \be m_\nu \to {\rm diag}[e^{i\alpha_e},
e^{i\alpha_\mu}, e^{i\alpha_\tau}]\cdot m_\nu \cdot {\rm
diag}[e^{i\alpha_e},e^{i\alpha_\mu}, e^{i\alpha_\tau}]. \ee
Apparently, the following combinations of matrix elements are
rephasing invariant: \be I_1 \equiv m_{e\mu}^2  m_{ee}^*
m_{\mu\mu}^*,  \ \ \, I_2 \equiv   m_{e\tau}^2  m_{ee}^*
m_{\tau\tau}^*,  \ \ \, I_3 \equiv m_{\mu\tau}^2  m_{\mu\mu}^*
m_{\tau\tau}^*, \label{rephasinginvariant} \ee so are the
equivalent combinations: $m_{\alpha \beta}^2/m_{\beta \beta}
m_{\alpha \alpha}$, where $\alpha\ne \beta$. Notice that under the
$\mu - \tau$ permutation, $I_3$ is invariant and \be I_1
\leftrightarrow I_2. \ee  Two other invariants read as \be I_4
\equiv \frac{m_{e\mu}^2 m_{\mu\mu}^*}{m_{e\tau}^2 m_{\tau\tau}^*}=
\frac{I_1}{I_2} \ee or $(m_{e\mu}m_{e\tau}^*)^2 m_{\tau\tau}
m_{\mu\mu}^* $
and \be I_5 \equiv \frac{m_{e\tau}m_{\mu\tau} m_{\tau
\tau}^*}{m_{e\mu} } = \sqrt{\frac{I_3 I_2}{I_1}}\ . \label{I5} \ee

Let us first formulate conditions for the complete
CP-conservation; {\it i.e.,} the case that the Dirac and Majorana
phases are both zero. For chiral states the CP transformation
coincides with C-conjugation: \be \nu_{\alpha} \rightarrow
\nu_{\alpha}^C \equiv C \bar{\nu}_{\alpha}^T , \ee where $C \equiv
i \gamma_2 \gamma_0 $. In terms of the charge conjugated states,
the mass terms (together with hermitian conjugate) can be written
as \be m_{\alpha \beta} \nu_{\alpha}^T C \nu_{\beta} + m_{\alpha
\beta}^* \nu_{\alpha}^{c T} C \nu_{\beta}^{c}. \ee Under CP
transformations $\nu_{\alpha} \rightarrow \nu_{\alpha}^c$ (an
additional phase can be removed by rephasing), and therefore
$m_{\alpha \beta} \to m_{\alpha \beta}^*$. Consequently, the
neutrino mass terms are CP invariant if \be m_{\alpha \beta} =
m_{\alpha \beta}^* , \ee that is, if the matrix elements can be
made real after rephasing.

Using invariants $I_i$ we can formulate the CP invariance in the
rephasing independent form. Similar analysis has been performed
recently in  \cite{recenthindu}, although, as we will see, some
differences between the two exist \footnote{Our results have been
presented in \cite{yasaman}.}. In particular, while
\cite{recenthindu} focuses on the textures compatible with
neutrino data our approach is more general and can find
application in contexts beyond the neutrino physics. Moreover, the
aim of \cite{recenthindu} is to formulate measures for
CP-violation rather than formulating criteria for CP-violation.

\begin{itemize}

\item
In the case that all the diagonal entrees of the mass matrix are
nonzero, CP is conserved if and only if the three invariants
$I_1$, $I_2$, $I_5$ are real:
\be {\rm Im} I_1 = 0, ~~~{\rm Im}
I_2 = 0, ~~~{\rm Im} I_5 = 0.
\label{3zero}
\ee
Notice that although $I_5$ can be written in terms of $I_1$, $I_2$
and $I_3$, one cannot replace the condition of Im$I_5=0$ with
Im$I_3=0$. This is illustrated by the following matrix \ba m_\nu
=\left[ \matrix{a & e & f\cr ... &b & id\cr ... & ... & c}\right]
\label{exception} \ea with real $a,e,f,b,c$ and $d$. This matrix
explicitly violates the CP but $I_1,I_2$ and $I_3$ associated with
it are all real.

\item
In the case that $m_{ee}=0$ but $m_{\mu \mu}$ and $m_{\tau\tau}$
are nonzero, CP is conserved if and only if the two invariants
$I_3$ and  $I_5$ are real: \be {\rm Im} I_3 = 0, ~~~{\rm Im} I_5 =
0. \label{inv-2} \ee

\item
In the case that $m_{ee}=m_{\mu \mu}=0$ but $m_{\tau \tau}$ is
nonzero, CP is conserved if and only if $I_5$ is real \be {\rm Im}
I_5 = 0. \label{inv-3} \ee

\item

Finally, if $m_{ee}=m_{\mu\mu}=m_{\tau \tau}=0$,  CP is conserved.

\end{itemize}

It is easy to check that conditions (\ref{3zero}, \ref{inv-2},
\ref{inv-3}) guarantee that the mass matrix can be made real by
rephasing and therefore the CP symmetry is completely conserved.

In order to test the violation of CP, one may choose to examine
another set of three independent combinations of $I_1$, $I_2$ and
$I_3$. So in this sense our criterion is not unique: depending on
the type of the texture, calculating a certain set of invariants
may have advantages over other sets. However, one must be aware of
the possibilities such as (\ref{exception}).

Combining $I_1$, $I_2$ and $I_3$, one can write new forms of
rephasing invariants:
$$J_1=m_{e \mu} m_{e \tau} m_{e e }^* m_{\mu \tau}^*=\sqrt{I_1 I_2
I_3^*}/\left( |m_{\mu \mu}|^2 |m_{\tau \tau}|^2\right)\ ; $$ $$
J_2=m_{e \mu} m_{\mu \tau} m_{e \tau }^* m_{\mu \mu}^*=\sqrt{I_1
I_3 I_2^*}/\left( |m_{ee}|^2 |m_{\tau \tau}|^2\right) \ ;$$
$$J_3=m_{e\tau}m_{\mu \tau} m_{e \mu}^* m_{\tau \tau}^*= \sqrt{I_2
I_3 I_1^*}/\left( |m_{ee}|^2|m_{\mu \mu}|^2 \right)\ .$$ In fact,
using (\ref{i-expr}) it is straightforward to show that ${\cal J}$
can be written  as the imaginary part of a combination of the
above invariants with real coefficient such as $|m_{\alpha
\beta}|^2$. As a result, the realness of the above rephasing
invariants guarantees that the Jarlskog invariant vanishes.
However, the opposite is not correct; {\it i.e,} we can have
${\cal J}=0$ but some of $I_i$ and $J_i$ may be complex. This is
due to the fact that, in contrast to the Jarlskog invariant, the
$I_i$ and $J_i$ contain information not only on the Dirac phase
but also on the Majorana phases.

\section{Zero Dirac CP-violating phase}

Due to  the fact that according to the neutrino oscillation data,
all the mass splitting, $\Delta m_{ij}^2\ (i\ne j)$, as well as
the values of $\sin 2 \theta_{12}$, $\sin 2\theta_{23}$ and $\cos
\theta_{13}$ are nonzero, the equality ${\cal J}=0$ implies $\sin
\theta_{13}\sin \delta=0$ [see Eqs. (\ref{calj},\ref{JCP})]. Below
we  derive the condition on the mass matrix that yield zero $\sin
\delta$ ({\it i.e.,} $\delta =k \pi$ with integer $k$) while
keeping $\sin \theta_{13}$ and Majorana phases  nonzero. The
possibility under consideration corresponds to a situation that
the forthcoming reactor and accelerator experiments  measure
$\theta_{13}$; however, the subsequent CP-violation searches would
report a null result. What can we learn from a sizeable $\sin
\theta_{13}$ but very small CP-violation, $J_{CP}\ll
s_{13}s_{12}c_{12}s_{23}c_{23}$?

Before deriving the necessary condition for $\delta = k \pi$, let
us formulate a criterion for zero $\sin \theta_{13}$. It is
straightforward to show that if $\sin \theta_{13}=0$,
$$
\tan 2 \theta_{23}={2|h_{\mu \tau}|\over h_{\tau \tau}-h_{\mu \mu}}
$$
and
$$
\tan \theta_{23}=\frac{|h_{e\tau}|}{|h_{e\mu}|},
$$ where $h$ is defined in
Eq.~(\ref{hmatr}). So the equality \be {2|h_{\mu \tau}|\over
h_{\tau\tau} - h_{\mu \mu}}={2|h_{e\tau} h_{e\mu}|\over
|h_{e\mu}|^2-|h_{e\tau}|^2} \label{criterion} \ee can be
considered as a test for $\sin \theta_{13}=0$. This criterion will
be useful to check whether zero ${\cal J}$ implies
$\sin \theta_{13}=0$ or $\sin \delta=0$.

Below we  consider  different situations that yield
vanishing $ \sin\delta$.

\subsection{Zero off-diagonal elements of $h_{\alpha \beta}$}

A trivial way to satisfy the condition  ${\cal J} = 0$  is to have
one or more vanishing off-diagonal elements of $h$: \be
h_{\alpha \beta} = 0, ~~  (\alpha \neq \beta). \label{offdiag} \ee
 This condition is both rephasing  and parametrization
invariant. In terms of the elements of the mass matrix, it implies
``orthogonality'' of the $\alpha$ and $\beta$ lines of the mass
matrix: \be \sum_i m_{\alpha i} m_{\beta i}^* = 0. \label{orthog}
\ee Apparently, these conditions are not necessary for  ${\cal J}
= 0$ and therefore lead to certain predictions for $\sin
\theta_{13}$. The hope is that a symmetry or principle will be
uncovered  that leads to the equality (\ref{offdiag}).

Let us consider three possibilities for different
$\alpha, \beta$ one by one.

1) $h_{e\mu}=0$. Using expression (\ref{hemu}) from the appendix
and taking into account the known experimental information that
1-2 and 2-3 mixings as well as $\Delta m^2_{ij}$ are nonzero we
find that $|h_{e\mu}|= 0$,  if \be \delta=0 \  \ {\rm and}  \ \
s_{13}s_{23} \Delta m^2_{31} + \Delta m^2_{21}
(s_{12}c_{12}c_{23}- s_{13}s_{23}s^2_{12}) =  0, \ee which implies
a rather small  value for the  1-3 mixing: \be \sin \theta_{13}
\approx - 0.5 \sin 2\theta_{12} \cot \theta_{23} r_{\Delta},
\label{13pred} \ee where \be r_{\Delta} \equiv \frac{\Delta
m_{21}^2}{\Delta m_{31}^2}. \ee Numerically we obtain $|\sin
\theta_{13}| \sim 0.016$. Unfortunately, such small values of
$s_{13}$ are beyond the reach of upcoming reactor experiments such
as Double CHOOZ \cite{chooz} and Daya Bay \cite{DayaBay} as well
as long baseline experiments \cite{T2K} (see, however,
\cite{nova}). Thus, a positive result in these experiments will
exclude such a possibility. Another solution is $\delta = \pi$ and
$\sin \theta_{13} \approx 0.5 \sin 2\theta_{12} \cot \theta_{23}
r_{\Delta}$.

So, if there is a  symmetry or principle that leads to $h_{e\mu} =
0$,  this equality  together with phenomenological input (nonzero
masses and two mixings) imply $\delta = 0$ and a value for $\sin
\theta_{13}$ given in Eq.~(\ref{13pred}). Inversely, confirming
relation (\ref{13pred})  will testify for such an underlying
symmetry. However it will not be sufficient  to  conclude that
$\sin\delta=0$. For $\sin \theta_{13}$ given in (\ref{13pred}),
$h_{e\mu}$ vanishes  only if  $\sin \delta $ also vanishes.


In terms of the elements of the mass matrix, the condition
$h_{e\mu}=0$  can be written as
\be
\label{allbuttautau}
m_{e\tau} = -\frac{1}{m_{\mu \tau}^*} (m_{ee}m_{e\mu}^* +
m_{e\mu}m_{\mu\mu}^*).
\ee
Although Eq.~(\ref{allbuttautau}) involves all the elements of the
mass matrix except $m_{\tau \tau}$ it does not, in general,  imply
a certain symmetry. However, as we show below, in some particular
cases, simple patterns of the mass matrix emerge.

For the normal hierarchy case, Eq.~(\ref{allbuttautau}) yields the
following form for the neutrino mass matrix \ba m_\nu = \left[
\matrix{ g & k & -k - [g k^*/A^*(1 + \epsilon_2^*)] + k
\epsilon_2^* \cr ... & A & A(1+\epsilon_2) \cr ... & ... & A
(1+\epsilon_3) } \right], \ea where $k, g  \ll A $ and $\epsilon_i
\ll 1$. Let us consider  the  special case of $g = 0$ and
$\epsilon_2 = 0$, so that the matrix reduces to \ba m_\nu\simeq
\left[ \matrix{ 0 & k & -k  \cr ... & A  & A \cr ... & ... & A
(1+\epsilon) } \right]. \label{matrixspe} \ea Apparently, the
above matrix has an approximate $\nu_{\mu} - \nu_{\tau}$ symmetry
broken in the 2-3 block. The parameters $k$ and $A$ can be made
real  by rephasing.

The matrix (\ref{matrixspe}) can be well motivated by a certain
symmetry. If $\epsilon = 0$,  the matrix  is  invariant under the
following transformations \ba \left[ \matrix{\nu_\mu \cr \nu_\tau}
\right] \Rightarrow e^{-i \theta} \left[\matrix{\cos \theta &
i\sin\theta \cr i\sin \theta & \cos \theta}\right] \left[
\matrix{\nu_\mu \cr \nu_\tau} \right], \label{trans} \ea \be
\label{etrans} \nu_e \Rightarrow e^{2i\theta}\nu_e \ee with
arbitrary $\theta$. Invariance under (\ref{etrans}) leads to a
vanishing $ee-$ element. The special case of  $\theta=\pi/2$
corresponds to  the $\nu_\mu \leftrightarrow \nu_\tau$ and $\nu_e
\to -\nu_e$ symmetry which implies $m_{e \mu}=-m_{e \tau}$ and
$m_{\mu \mu}=m_{\tau\tau}$. Finally, the invariance  under general
transformations (\ref{trans}) leads to $m_{\mu\mu}=m_{\mu \tau}$.
Notice that  the equality of $m_{\mu \mu}$ and $m_{\mu \tau}$ does
not follow from the $\mu-\tau$ permutation symmetry.

The symmetry under  (\ref{trans},\ref{etrans}) does not  explain
the hierarchy $k/A\sim 0.1$. This can be either  accidental
(notice that the hierarchy is not too large) or a consequence of
 approximate $L_e$ conservation. Nonzero $\epsilon$ (which is
necessary for explaining the neutrino data) breaks the symmetry
under (\ref{trans}). Notice that in the charged lepton sector (as
well as in the quark sector) the symmetry is broken by a similar
structure; {\it i.e.,}  $m_\tau \gg m_\mu$. The two can be
related. In this connection one can consider  two different
contributions to the neutrino mass matrix such that while the
dominant contribution obeys the symmetry under
(\ref{trans},\ref{etrans}), the subdominant contribution, which
violates the symmetry, has a hierarchical structure similar to the
one in the charged lepton
sector. \\

Let us consider the phenomenological consequences of  matrix
(\ref{matrixspe}). Diagonalization leads to the mixing angles \be
\tan 2 \theta_{12} = {4 \sqrt{2} k\over |\epsilon| A}, \ \ \ \sin
\theta_{13}= -{k |\epsilon| \over 4 \sqrt{2} A} =-
\frac{|\epsilon|^2}{32}\tan 2 \theta_{12}, \label{phenomexp} \ee
and $\delta = 0$, which is expected because  $\theta_{13}$  is
nonzero. Moreover,
$$\theta_{23}=
\frac{\pi}{4}-\frac{{\rm
Re}[\epsilon]}{4}-\frac{|\epsilon|^2}{8}.$$ 
For the mass eigenvalues we obtain \be \label{eigen}
|m_{1,2}|^2=2k^2+\frac{A^2}{8}|\epsilon|^2 \pm
\sqrt{\left(2k^2+\frac{A^2}{8}|\epsilon|^2\right)^2 -4 k^4}, ~~~
|m_3|^2 \approx 4A^2, \ee thus \be
 r_{\Delta}
= \frac{|\epsilon|^2}{16 \cos 2\theta_{12}}. \ee If $\epsilon$ is
real, we obtain \be \tan 2\theta_{23} = \frac{2}{\epsilon}\ee
that results in the following relations among the observables:
\be \tan^2 2 \theta_{23} = \frac{1}{4 r_{\Delta}\cos
2\theta_{12}}. \label{predi} \ee From (\ref{predi}) we obtain a
deviation from the  maximal mixing \be \frac{1}{2} - \sin^2
\theta_{23} = \left[ \frac{1}{r_{\Delta}\cos 2\theta_{12}} + 4
\right]^{-1/2} \approx 0.1 \ . \ee
Rewriting $\sin \theta_{13}$ in (\ref{phenomexp}) in terms of the mixing
parameter, we find \be \sin \theta_{13} =- \frac{\tan
2\theta_{12}}{8\tan^2 2 \theta_{23}} = - \frac{1}{2} \sin
2\theta_{12}r_\Delta,  \label{38} \ee
that coincides with
(\ref{13pred}) for $\cot \theta_{23} \approx 1$.

Establishing  small ($\sim r_{\Delta}$) 1-3 mixing and a
relatively large deviation of 2-3 mixing from maximal: $\sin^2
2\theta_{23} \approx 0.96$,  will be in support of the considered
possibility. Such a deviation from maximal mixing can be tested by
future   long baseline experiments \cite{T2K,nova}.
%
These are typical features of a mass matrix whose $\nu_{\mu} -
\nu_{\tau}$ symmetry is broken in the $\nu_{\mu} - \nu_{\tau}$
block. \vskip 0.3 cm

If $\epsilon$ is complex, $\epsilon = |\epsilon|
exp(i\phi_{\epsilon})$, the Majorana CP-violating phases will be
nonzero. Let us evaluate these phases. According to our convention
$m_1$ is real, thus from the condition $(m_\nu)_{ee} = 0$ we
obtain
\be {\rm Im}(m_2) s_{12}^2+{\rm Im}(m_3) s_{13}^2=0, \ee or using
Eqs. (\ref{phenomexp},\ref{eigen}) \be \sin \phi_2 \approx \sin
\phi_3 (r_{\Delta})^{3/2}. \ee Thus, in the first approximation
the phase of $m_2$ can be neglected. To obtain the phase of $m_3$,
we use the rephasing invariant $I_{3}$
$$
{m_{\tau \tau} m_{\mu \mu} \over m_{\mu \tau}^2} =1+\epsilon .
$$
Since the 1-3 mixing is small, the $I_3$ invariant can be
rewritten as \be {(\bar{m}_2 c_{23}^2 + m_3 s_{23}^2)( \bar{m}_2
s_{23}^2 + m_3 c_{23}^2 )
 \over  (\bar{m}_2 - m_3)^2 s_{23}^2 c_{23}^2}
 = 1+ \epsilon,
\label{inv3rel}
\ee
where
$$
\bar{m}_2 \approx m_2 c_{12}^2 + m_1 s_{12}^2.
$$
Notice that $\bar{m}_2$ is real in our approximation. From
(\ref{inv3rel}) we obtain \be {\bar{m}_2 m_3
 \over  (\bar{m}_2 - m_3)^2 s_{23}^2 c_{23}^2} = \epsilon,
 \ee and since $\bar{m}_2 \ll m_3$
$$
\phi_3 \simeq -\phi_\epsilon.
$$
So, the phase of $m_{3}$ is equal to the phase of $\epsilon$
and in general can be large.\\


2) $h_{\tau e}=0$: This equality  results in zero $\delta$ and \be
\sin \theta_{13} \approx 0.5 \sin 2\theta_{12} \tan \theta_{23}
r_{\Delta}, \label{another13} \ee or $\delta=\pi$ and $\sin
\theta_{13} = - 0.5 \sin 2\theta_{12} \tan \theta_{23}
r_{\Delta}$. The equality $h_{\tau e}=0$ means that \be m_{e
\mu}=-{m_{ee}m_{e \tau}^*+m_{e \tau}m_{\tau \tau}^* \over m_{\mu
\tau}^*} , \ee and in the case of normal hierarchical scheme this
leads to the mass matrix
\begin{eqnarray}
m_{\nu} =
\left[ \matrix{ g' & -\frac{g' k'^*+k' A^*}{A^* (1+\epsilon_2'^*)}
& k' \cr ... & A(1+\epsilon_3') & A(1+\epsilon_2') \cr ...  & ...
& A} \right].
\end{eqnarray}
If   $g' = \epsilon_2' = 0$, we obtain
\begin{eqnarray} m_\nu=\left[ \matrix{ 0 & -k' & k'
\cr ...  & A(1+\epsilon') & A\cr ... &... &A}\right].
\end{eqnarray}
The mixing angles and the splittings are  the same as in the
previous case  except that $$\theta_{23}=\frac{\pi}{4}+\frac{{\rm
Re}[\epsilon']}{4}+\frac{|\epsilon'|^2}{8}.$$

3) Equality $h_{\mu \tau}=0$ leads to
$$\Delta m_{31}^2 s_{23}c_{23}+\Delta m_{21}^2 (s_{23}c_{23}\cos
2 \theta_{12}-s_{12}c_{12}s_{13}\cos 2 \theta_{23}) =0
$$
which is not compatible with the data.

In summary, in this subsection, we have  found that although
$h_{\mu \tau}=0$ is not compatible with the data, $h_{e \mu}=0$ or
$h_{e \tau}=0$ yield vanishing $\sin \delta$  and a small but
nonzero value for $s_{13}$ of order of $\Delta m_{21}^2/\Delta
m_{31}^2$.
\subsection{Small Dirac phase}

According to (\ref{i-expr}) if all the  elements of $h_{\alpha
\beta}$ are real, no  CP-violating effects  appear in
oscillations but in general  $\theta_{13}$ and the Majorana phases
can be nonzero. By straightforward but cumbersome calculations, it
can be shown that
to the leading order in $\sin \theta_{13}$ and $\cos
2\theta_{23}$, the mass matrix which satisfies this condition can
be parameterized as
 \begin{eqnarray}
 m_\nu=\left[\matrix{r-2xs-t & s(1+\eta x-\alpha)+\eta t&
 -s(1-\eta x+\alpha)+\eta t \cr ...  & r+2\alpha t-s\eta &t\cr
 ...  & ... & r-2\alpha t+\eta s}\right],
\label{real-off-diag}
 \end{eqnarray}
where $\alpha,\eta \ll x\sim 1 $ are real numbers but $r,$ $s$ and
$t$ can in general be  complex quantities.
  It can be shown that regardless of the mass scheme
(hierarchical or degenerate; normal or  inverted) the parameters of this  mass
matrix are immediately related to the observables as
$$\cot 2 \theta_{12}=\frac{x}{\sqrt{2}},
\ \ \ \ \ \ c_{23}= (1-\alpha)/\sqrt{2}$$  and
$$ \sin \theta_{13} = \eta/\sqrt{2}.
$$
In order to reproduce  the observed neutrino mass splitting, the
complex parameters $r$, $s$ and $t$ should satisfy certain  relations.
It can be shown that \begin{eqnarray}m_1&=&{-2 (r-t)
s_{12}^2-2xsc_{12}^2 \over \cos 2\theta_{12}} \cr m_2&=&{2 (r-t)
c_{12}^2+2xss_{12}^2 \over \cos 2\theta_{12}} \cr m_3 &=&
r+t.\label{complexmasses}
\end{eqnarray}
Thus, $$\Delta m_{21}^2= {4\left( |r-t|^2-|s|^2 x^2\right) \over
\cos 2\theta_{12}}.$$ The normal mass hierarchy requires
$|r-t|^2,x^2 |s|^2\ll |r+t|^2$;  while, in order to achieve
inverted hierarchy, one needs $|r+t|^2, (|r-t|^2-x^2|s|^2) \ll
(|r-t|^2+x^2|s|^2)$.

 For the mass matrix given in Eq.~(\ref{real-off-diag}) with
general but small $\alpha$ and $\eta$ the Dirac phase is small but
 nonzero: $\delta \sim {\cal O}({\rm Max}[\eta,\alpha]) \ll
1$. In the specific case that \be \label{tinytiny} \eta^2+2\eta
\alpha x-2 \alpha^2=0,\ee the Dirac phase vanishes, exactly. This
can be verified in two steps: 1) If the above relation holds,
$h_{e\mu}$, $h_{e \tau}$ and $h_{\mu \tau}$ are all real which
means the Jarlskog invariant is zero; 2) using the criterion
(\ref{criterion}), we find that $s_{13}\ne 0$ which means $\sin
\delta$ must vanish.

 The cases discussed in the previous section
($h_{e\mu},h_{e\tau}=0$) can be considered as special instances of
the mass matrix with the form in Eq.~(\ref{real-off-diag}),
provided that (\ref{tinytiny}) is satisfied. For example, \be
\label{thesame}\eta\simeq \pm \frac{4}{x}{ |r-t|^2-x^2|s|^2 \over
|r+t|^2-4\left|(r-t)s_{12}^2+xs
c_{12}^2\right|^2/\cos^22\theta_{12}} \ee yields vanishing
$h_{e\mu}$. Using (\ref{complexmasses}) and remembering that
$\eta=\sqrt{2}\sin \theta_{13}$, it can be confirmed that
(\ref{thesame}) yields  the same relation between $\sin
\theta_{13}$ and the mass splittings that we expected in the case
of $h_{e \mu}=0$; {\it i.e.,} Eq.~(\ref{thesame}) corresponds to
(\ref{38}).

In what follows we show that by a change of basis, matrix
(\ref{real-off-diag}) acquires a simple form which will be easier
to incorporate in models. Let us define $(\tilde{\nu}_0 \
\tilde{\nu}_+ \ \tilde{\nu}_-)\equiv (\nu_e \ \nu_\mu \ \nu_\tau)
V_b^T, $ where $V_b$ is a unitary matrix which neglecting {\cal
O}$(\eta^2,\alpha^2)$, can be written as \ba V_b =\left[\matrix{ i
& -i \eta/2 & -i \eta/2 \cr \eta/\sqrt{2} & (1+\alpha)/\sqrt{2} &
(1-\alpha)/\sqrt{2} \cr 0 & i(1-\alpha)/\sqrt{2} &
-i(1+\alpha)/\sqrt{2} } \right]. \label{newbasis} \ea It is
straightforward to show that  in this basis, up to {\cal
O}$(\eta^2,\alpha^2)$, matrix (\ref{real-off-diag}) obtains the
following simple form
\ba \label{simpleinnew}\left[ \matrix{-r+t+2xs & 0 & -\sqrt{2} s
\cr 0 & r+t & 0 \cr -\sqrt{2}s & 0 & -r+t}\right]. \ea

In this subsection, we have derived the general form of a neutrino
mass matrix for which $|\sin \delta| \ll 1$. We have then shown
that by changing the basis, this matrix acquires a simple form
which will be easy to incorporate into models.
\subsection{Zero Dirac phase with nonzero
elements of $h_{\alpha \beta}$}

Consider a matrix $h$ with  vanishing  Jarlskog invariant but
non-zero off-diagonal elements.
In general the equality $J_{CP}=0$ is equivalent to the condition
that by rephasing the neutrino fields the $h$ matrix can be made
real. However,  because of the possibility of nonzero Majorana
phases, $J_{CP}=0$ does not necessarily mean that $m_\nu$ can be
made real by rephasing. In this section we formulate the necessary
conditions  on $m_\nu$ for $J_{CP}=0$ in a specific class of mass
patterns, demonstrating how the invariants defined in the previous
section can simplify the analysis.

Let us consider a matrix for which $I_3$, $I_4$ and $I_5$ are all
real. After a proper rephasing, the general form of such a matrix
can be written as  \ba m_\nu=\left[ \matrix{ u & v & nv \cr ... &
B & lB \cr ... & ... & (1 + \kappa) B }\right], \label{matrix3}
\ea where $l$, $n$ and $\kappa$ are real numbers but $u$, $v$ and
$B$ can have complex values. This assumption means that  the
phases of $m_{e\mu}$ and $m_{e\tau}$ are equal, and also all the
elements of $\mu - \tau$ block have the same phase. In this case
by rephasing the neutrino fields one can eliminate the phases from
all the elements but $m_{ee}$; {\it i.e.,} only $u$ remains
complex. Apparently, if $m_{ee} = 0$, there will be no
CP-violation in the lepton sector.


Matrix (\ref{matrix3}) automatically  gives
$$
{\rm Im}[h_{\mu\tau}] = 0.
$$
The two other elements are equal to
$$
h_{e\mu} = u v^*+vB^*(1+ ln),
$$
$$
h_{\tau e} = nv u^*+B v^*[l+ n (1 + \kappa)].
$$
Then the condition of the absence of the CP violation in
oscillations, ${\cal J}=0$, reduces to \be {\rm Im}[u B(v^*)^2] (l
+ n(1 + \kappa)) - n(1 + ln)] = 0. \label{zeroI} \ee Let us
consider solutions of this equation and their implications.

1) The condition (\ref{zeroI}) is satisfied if $u = 0$, as we
have  noticed before. Two other trivial solutions, $B = 0$ and $v
= 0$ are ruled out by the neutrino data.

2) Another possible solution (again for arbitrary $l$, $n$ and
$\kappa$) is \be {\rm Im}[I_1^*]={\rm Im}[u B(v^*)^2] = 0.
\label{inv22} \ee  For two other invariants we obtain the
following: $I_2 = n (1 + \kappa) v^2 u^* B^*$  - that  has the
same imaginary part as $I_1$,  and $I_5 = ln (1 + \kappa) |B|^2$
which is real. Therefore if condition (\ref{inv22}) is satisfied,
all the invariants, $I_1, I_2, I_5$ will be real and, according to
the general consideration of sec. 2, there will be no CP-violation
even for nonzero $u$; {\it i.e.,} all the  physical phases will
vanish.

The same conclusion can be obtained in a different way. As we have
already pointed out, we can make  all the parameters of matrix
(\ref{matrix3})
 except $u$ real by rephasing. In this case,  condition
(\ref{inv22}) means that $u$ should be real too. Notice that the
1-3 mixing is,  in general,  nonzero.

3) The last non-trivial solution of (\ref{zeroI}) is \be l + n(1 +
\kappa) =   n(1 + ln) .\label{conds13} \ee Performing
diagonalization of the mass matrix and remembering that the 1-2
mixing is nonzero, it is straightforward to check that the
condition (\ref{conds13}) leads to a vanishing 1-3 mixing. So,
${\cal J} = 0$ is satisfied trivially due to the zero mixing.

Thus,  with a matrix of form  (\ref{matrix3}), the Jarlskog
invariant can vanish if and only if either the lepton sector is
CP-invariant ({\it i.e.,} both the Dirac and Majorana phases
vanish) or $\sin \theta_{13}=0$. In other words, for
(\ref{matrix3}) we cannot have $\sin \theta_{13}\ne 0$, $\sin
\delta=0$.
\section{Maximal CP violating phase}

Several neutrino mass matrix textures have been proposed that
predict a  maximal value for the  Dirac CP-violating  phase
\cite{A4CP,yasue,mohapatra,xing}. In other words,  for given
values of mixing angles, they predict the Jarlskog invariant to be
maximal which implies $|\sin \delta|=1$. For the experiments
proposed to directly search for the  CP-violation in the neutrino
oscillations, this means that for given values of mixing angles,
the asymmetry $P(\nu_\mu \to \nu_e)-P(\bar\nu_\mu \to \bar\nu_e)$
is maximal.

 It has been suggested in \cite{triangle}
to use astrophysical neutrinos to determine the value of $\delta$
without directly measuring the CP-violating effects. For stable
neutrinos, the effect of $\delta$ will be too small to resolve.
However, if at cosmological distances the heavier neutrinos decay
into $\nu_1$, the ratio of neutrino fluxes of different flavors
will be  sensitive to $\sin \theta_{13}\cos\delta$: $$
\Phi_{\nu_e}:\Phi_{\nu_\mu}:\Phi_{\nu_\tau}=|U_{e1}|^2:|U_{\mu 1
}|^2:|U_{\tau 1}|^2,$$ which for $c_{23}^2=s_{23}^2=1/2$ and
$s_{13}\ll 1$ can be written as
$$ |U_{e1}|^2:|U_{\mu 1}|^2:|U_{\tau 1}|^2\simeq
c_{12}^2:{s_{12}^2+2s_{12}c_{12}s_{13}\cos\delta\over
2}:{s_{12}^2-2s_{12}c_{12}s_{13}\cos\delta\over 2}. $$ In the case
of the maximal CP-violating phase, like in the case of $s_{13}=0$,
we expect $\Phi_{\nu_\mu}/\Phi_{\nu_e}=\tan^2\theta_{12}/2$, which
can be in principle  checked by Icecube
\cite{triangle,flavorratio}. Thus, from the point of view of the
indirect measurements, the maximal CP-violation value of the phase
($\cos \delta=0$) is also special.

As noticed in \cite{scott,grimus}, the maximal Dirac CP-violating
phase can be explained by the so-called $\mu  - \tau$ reflection
symmetry under which $\nu_\mu$ and $\nu_\tau$ transform into the
CP-conjugate of each other. We first study various  new aspects of
this symmetry, and then in section~\ref{nonzerosection},
generalize it to accommodate non-maximal values of $\delta$ as
well as a deviation of the 2-3 mixing from $\pi/4$. 

\subsection{$\mu -  \tau$ reflection symmetry \label{cpreflection}}


The $\mu\ - \ \tau$ reflection \cite{scott,grimus} is defined as
follows \be \nu_e \to  \xi_1 \nu_e^c, ~~~ \nu_\mu \to \xi_2
\nu_\tau^c, ~~~ \nu_\tau \to  \xi_3 \nu_\mu^c, \label{defsym} \ee
where $\xi_i$ are phase factors. These transformations  are
combinations of  charge conjugation, parity and permutation in the
flavor space. 
Notice that if $\xi_2 \ne \xi_3$,  $m_{e\mu}$ and $m_{e \tau}$
should both vanish which contradicts the observations. Thus, we
should take \be \xi_2=\xi_3. \label{phasesrel} \ee The most
general form of neutrino mass matrix invariant under the $\mu\ - \
\tau$ reflection transformations is  \ba m_\nu=\left[ \matrix{i
f\xi_1^* & w e^{-i \sigma} & -w \xi_1^*\xi_2^* e^{i \sigma} \cr
... & ye^{2i\beta} & -iz\xi_2^* \cr ... & ... & -y
e^{-2i\beta}(\xi_2^*)^2}\right], \label{reflection...} \ea where
$w,y,z,f$  and $\sigma$ are all real. By rephasing $\nu_\mu \to
e^{-i \beta} \nu_\mu$, $\nu_\tau \to i\xi_2 e^{i \beta} \nu_\tau$,
$\nu_e\to \sqrt{-i\xi_1}\nu_e$ and redefining $\sigma$, the mass
matrix obtains the form~\footnote{A similar texture has been
considered in \cite{yasue,mohapatra,xing} where the dominant part
of the mass matrix is taken to be symmetric under the $\mu -\tau$
exchange and the nonzero phases are introduced as parameters that
break the $\mu-\tau$ symmetry. In contrast, the $\mu -\tau$
reflection symmetry, even if exact, can accommodate nonzero
phases.}
\ba m_\nu=\left[ \matrix{f & we^{-i\sigma} & -w e^{i \sigma} \cr
... & y & z \cr ... & ... & y } \right]. \label{matrixmax} \ea For
such a mass matrix, ${\cal J}$ [see Eq.~(\ref{calj})] is equal to
$$
{\cal J} =    -2w^2y \sin2 \sigma\left[
w^2(-z+f)+z(f-z)^2-y^2z+yw^2\cos2\sigma \right]
$$ which, in general,
is nonzero for  $\sin 2\sigma, w, y \neq 0$. In the following we
show that for $\sin2\sigma \neq 0$ the matrix (\ref{matrixmax})
implies a nonzero 1-3 mixing and $\delta = \pi/2$.

Eq.~(\ref{matrixmax}) yields the following relations among the
elements of the  matrix $h$: \be h_{\mu\mu} = h_{\tau\tau} ~~~[ =
w^2+y^2+z^2], \label{hheq1} \ee and \be h_{e\mu} =- h_{\tau e} =-
h_{e \tau}^*~~~[= w(f e^{i \sigma}-z e^{i \sigma}+ y e^{-i
\sigma})]. \label{hheq2} \ee Using explicit expressions for
$h_{\alpha\beta}$ in terms of  the oscillation parameters we find
from
 Eq.~(\ref{hheq1})
\be \label{theta23} \cos 2\theta_{23}
= -4\rho_{\Delta} s_{12}c_{12}s_{23}c_{23}s_{13} \cos \delta, \ee
where
$$ \rho_\Delta={\Delta m_{21}^2 \over \Delta m_{31}^2
c_{13}^2+(s_{12}^2s_{13}^2-c_{12}^2)\Delta m_{21}^2}.
$$
From Eq.~(\ref{hheq2}),  $ |h_{e\mu}| = |h_{\tau e}|$, we obtain
$$
\cos 2\theta_{23} \left[ (\Delta m_{31}^2)^2 s_{13}^2
c_{13}^2-(\Delta m_{21}^2)^2 s_{12}^2 c_{12}^2 c_{13}^2\right]
$$
\be \label{pair} -4 \Delta m_{31}^2 \Delta m_{21}^2 s_{13}c_{13}^2
s_{23}c_{23} s_{12}c_{12}\cos \delta = 0. \ee For the observed
values of mass squared differences and mixing angles,  equalities
(\ref{theta23}) and (\ref{pair}) are satisfied if $\cos 2
\theta_{23} = 0$ ({\it i.e.,} 2-3 mixing is maximal) and \be \sin
\theta_{13}\cos\delta=0.\label{maxzero} \ee For $\sin
\theta_{13}=0$ and $\cos 2\theta_{23} = 0$ the  rephasing
invariants $I_1$ and $I_2$  (\ref{rephasinginvariant}) are equal,
which in the case of (\ref{matrixmax}), is realized provided that
$e^{4i\sigma}=1$. Notice that for $\sin \theta_{13}=0$, or in
other words for $\sigma=0,\pi,\pm \pi/2$, the $\mu-\tau$ symmetry
\cite{takeshi} is restored and the CP in oscillations is
conserved. For a general value of  $\sigma$ (not belonging to
$\{0,\pm\pi/2,\pi\}$), $\sin\theta_{13}$ is nonzero and
Eq.~(\ref{maxzero}) can be satisfied only if $\cos \delta=0$.

 In addition to predicting maximal  values for $\theta_{23}$ and $\delta$, the
 $\mu -\tau$ reflection symmetry
 predicts zero (or equal to $\pi$) Majorana phases. This can be
proved by explicitly  diagonalizing  the matrix $m_{\nu}$. This conclusion
holds for any mass  scheme: normal/inverted,
hierarchical/degenerate.

The $\mu- \tau$ reflection symmetry is strongly broken in the
charged lepton sector. As a result,
we expect the radiative corrections induced by the charged lepton
sector to break this symmetry in the neutrino sector, too (see  sec.
\ref{RGE}).

In what follows, we study the normal mass  hierarchy  in
the context of the $\mu-\tau$ reflection symmetry. In order to
reproduce the normal hierarchy, the elements of
(\ref{matrixmax}) should satisfy:
 $$
  | y^2-z^2|\sim w^2,f^2\ll y^2,z^2.
$$
Using these inequalities it is straightforward to show that \be
\tan \theta_{23}=1,\ \ \tan 2\theta_{12}\simeq {2\sqrt{2}w \cos
\sigma \over y-z - f}, \ \  \sin \theta_{13}\simeq \frac{\sqrt{2}
w \sin\sigma }{ y+z}, \ee and \be m_3=y+z, \ \ m_{1,2}={y-z-f\pm
\sqrt{(z-y-f)^2+8w^2\cos^2\sigma} \over 2} \ee  and finally, as
expected, $\delta=\pi/2$ but the Majorana phases vanish. Combining
the above formulas and the information on masses and mixing we
obtain  $w\cos\sigma /(y+z)\sim 0.1$. For $\tan \sigma \sim 1$,
$\theta_{13}$ can saturate its  upper bound. Inversely, the bound
on $\sin \theta_{13}$ can be interpreted as an upper bound on
$\tan \sigma$. Notice that for $\sin \sigma \to 0 $, the
$\mu-\tau$ exchange symmetry is restored and as a result $\sin
\theta_{13}=0$, and consequently,  there will be no CP-violating
effects.

Now let us consider the possibility to reproduce  matrix
(\ref{matrixmax}) for the normal mass hierarchy by the seesaw
mechanism. Without loss of generality we choose the  mass basis
for the right-handed neutrinos. Suppose that under the $\mu-\tau$
reflection symmetry, the right-handed neutrinos transform into the
charge conjugates of themselves, $N_i \to N_i^c$. In order  for
the right-handed neutrino mass matrix to be symmetric under the
$\mu-\tau$ reflection, it should be real:
$$  {\cal L}_{{\rm mass}}=-\sum_{i=1}^3 M_i N_i^TCN_i   \ \ \ {\rm
with }  \ \ \ M_i=M_i^*.$$ Moreover, the $\mu-\tau$ reflection
symmetry with $N_i \to N_i^c$ implies the following form for the
neutrino Yukawa couplings
 \ba Y_\nu = \left[ \matrix{a_1 & b_1 e^{i \kappa_1}&
b_1 e^{-i \kappa_1} \cr a_2 & b_2 e^{i \kappa_2}& b_2 e^{-i
\kappa_2} \cr a_3 & b_3 e^{i \kappa_3}& b_3 e^{-i \kappa_3}
}\right] \label{yukkk} \ea with real $a_i$ and $b_i$.

 We assume that
the right-handed neutrino $N_3$ dominates in generation of the
light mass matrix (single right-handed neutrino
dominance~\cite{king}). This implies $a_1^2,|b_1|^2,|b_2|^2\ll
|b_3|^2 M_1/M_3$. Moreover, suppose the dominant  part of the
Lagrangian that  involves $N_3$ preserves $L_e$; {\it i.e.,}
$a_3=0$. Then, the mass matrix of the light neutrinos will be
equal to
 \ba
v^2\left[\matrix{ \frac{a_1^2+(a_1^*)^2}{M_1} & \frac{a_1
b_1+a_1^* b_2}{M_1} & \frac{a_1b_2^*+a_1^*b_1^*}{M_1} \cr ...  &
\frac{b_3^2}{M_3} + \frac{b_2^2+b_1^2}{M_1} &
\frac{|b_3|^2}{M_3}+\frac{b_2b_1^*+b_1b_2^*}{M_1} \cr ...  & ... &
\frac{(b_3^*)^2}{M_3} +  \frac{(b_2^*)^2 +(b_1^*)^2}{M_1} }\right]
\ea which, considering the realness of $M_i$, is precisely of
form (\ref{matrixmax}).

\subsection{$\mu - \tau$ exchange  versus $\mu-\tau$ reflection symmetry}

Let us  compare the  phenomenological consequences of the
$\mu-\tau$ reflection and exchange symmetries and discuss the
necessary and sufficient conditions for having maximal
CP-violating phases. As follows from the discussion in sec.
(\ref{cpreflection})  both symmetries
guarantee that \be h_{\mu \mu}= h_{\tau \tau}, \ \ \ \ \  \ |h_{e
\mu}|=|h_{e \tau}|\label{twoeq}. \ee As shown in section
(\ref{cpreflection}), the above  equalities, in turn, lead to
$\cos 2\theta_{23} = 0$ (maximal 2-3 mixing) and $\sin
\theta_{13}\cos\delta = 0$. However, realization of the last
equality is different for the two symmetries: while $\nu_\mu
\leftrightarrow \nu_\tau$ implies $\sin \theta_{13}=0$, nonzero
$\sin \theta_{13}$ is compatible with the $\mu-\tau$ reflection
symmetry,  and as a result,  the latter symmetry requires that
$\cos \delta=0$.

Now, let us discuss whether equalities $\cos 2\theta_{23} = \sin
\theta_{13}\cos\delta=0$ necessarily imply the aforementioned
symmetries.
 It is well-known (and in fact,  straightforward to confirm)
 that the equalities
\be \cos 2 \theta_{23}= \sin \theta_{13}=0 \label{compatible} \ee
are compatible with neutrino data ({\it i.e.,} nonzero $\sin 2
\theta_{12}$ and $\Delta m_{21}^2$) only if there is a $\mu-\tau$
symmetry. In other words, (\ref{compatible}) implies $I_1=I_2$
[see Eq.~(\ref{rephasinginvariant})] and
$|m_{\tau\tau}|=|m_{\mu\mu}|$. On the contrary, to have \be \cos 2
\theta_{23}=\cos \delta=0 \label{reflectiongives} \ee the
$\mu-\tau$ reflection symmetry is not a necessary condition.
Indeed,  the two conditions $h_{\mu \mu}= h_{\tau \tau}$ and $| h_{e
\mu}|=|h_{e \tau}|$   can be simultaneously satisfied while
$|m_{e\mu}| \neq |m_{e\tau}|$.
To show this, let us   write the elements
of  $m_\nu$ in terms of oscillation parameters:
$$m_{e
\mu}=s_{12}c_{12}c_{23}c_{13}(|m_2|e^{i\phi_2}-|m_1|)$$
$$+s_{23}s_{13}c_{13}(|m_3| e^{i \phi_3}e^{-i\delta}-|m_1| c_{12}^2
e^{i\phi_1}e^{i \delta}-|m_2|s_{12}^2 e^{i \phi_2}e^{i\delta}),$$
$$m_{e\tau}=-s_{12}c_{12}s_{23}c_{13}(|m_2|e^{i\phi_2}-|m_1|e^{i
\phi_1})$$
$$+c_{23}s_{13}c_{13}(|m_3| e^{i \phi_3}e^{-i\delta}-|m_1| c_{12}^2
e^{i\phi_1}e^{i \delta}-|m_2|s_{12}^2 e^{i \phi_2}e^{i\delta}).
$$
Apparently for $|m_{e\mu}|=|m_{e \tau}|$, in addition to
equalities (\ref{compatible}), the sines of the Majorana phases
must also vanish. However, for general nonzero phases,
$\left(|m_{e \mu}|-|m_{e \tau}|\right)/\left(|m_{e \mu}|+ |m_{e
\tau}|\right)\sim 1$.  The maximal 2-3 mixing and Dirac phase do
not guarantee that the $\mu-\tau$ reflection symmetry exists. On
the other hand,  the $\mu-\tau$ reflection symmetry is not
necessary for maximal CP-violating phase.

Summarizing, zero Majorana phases and Eq.~(\ref{reflectiongives})
imply $I_1=I_2^*\ne I_2$ and $I_3=I_3^*$ which (by appropriate
rephasing) results in matrix (\ref{matrixmax}). However, if
Majorana phases are nonzero even for (\ref{reflectiongives}), the
element $|m_{e\mu }|$ can be different from $|m_{e \tau}|$.
Therefore,  to prove the existence of a  $\mu-\tau$ reflection
symmetry not only confirming (\ref{reflectiongives}) is necessary
but one  has to also show that the Majorana phases vanish.

Notice that the phases of off-diagonal elements of $h$ are not
rephasing invariant and are not therefore physical. As a result,
despite the claim in \cite{yasue}, $h_{e \tau}=-{\rm sgn[s_{23}]}
h_{e \mu}^*$ does not guarantee that $\delta=\pi/2$, $\cos
\theta_{23}=1$. In fact,  for any pair of ($\delta, \theta_{23})$
satisfying Eq.~(\ref{pair}), we can have $|h_{e\mu}|=|h_{e\tau}|$.
Then, by appropriate rephasing of the fields the condition $h_{e
\tau}=-{\rm sgn[s_{23}]} h_{e \mu}^*$ can be satisfied.

\subsection{Renormalization group  effects\label{RGE}}

The CP-violating phase is not invariant under renormalization.
Furthermore, the physics responsible for a certain pattern of the
mass matrix  can become manifest only at very high scales, {\it
e.g.,} the Grand unification scale $M_{GU}$. The RG effects should
be therefore taken into account when confronting with the low
energy observations.

The radiative corrections change the matrix and break the
$\mu-\tau$ reflection symmetry, so the value of $\delta$ is
expected to be modified. For illustration let us consider the
running of the effective $D = 5$ operator
$$
|H|^2\sum_{ij}\nu_i^T \frac{(m_\nu)_{ij}}{\langle H\rangle^2}
C\nu_j
$$
from the scale of decoupling of the heavy neutrinos (the seesaw
scale) down to the low energies. Neglecting the corrections
proportional to $m_{ \mu}^2/\langle H\rangle^2$ and
$m_{e}^2/\langle H\rangle^2$ and absorbing the flavor-independent
corrections in the definition of the overall mass scale we find
that matrix (\ref{matrixmax}) will modify into
\ba m_\nu=\left[ \matrix{f & we^{-i\sigma} & -w e^{i \sigma}(1+
\tilde{\epsilon}) \cr ... & y & z (1+\tilde{\epsilon})\cr ... &
... & y(1+2\tilde{\epsilon}) } \right]. \ea In the standard model,
 we have $ \tilde{\epsilon} \sim
\frac{\log[\Lambda/M_Z]}{16\pi^2}\frac{m_\tau^2}{\langle
H\rangle^2}$ , which setting the cutoff  $\Lambda= 10^{12}$ GeV,
is of order of $10^{-5} $.  In the supersymmetric version with
large $\tan \beta$ the corrections can be much larger:
$\tilde{\epsilon} \sim 10^{-2}$.

As a result of the corrections, both $\theta_{23}$ and $\delta$
are shifted from their maximal mixing and CP-violation values
by a tiny amount. It
can be shown that
$$
\cos\theta_{23}-\frac{1}{\sqrt{2}}={ 2y^2+w^2\over 4\sqrt{2} yz}
\tilde{\epsilon},
$$
and for  normal mass hierarchy ($w^2\ll y^2\simeq z^2$) the
deviation from maximal mixing is approximately equal to
$\tilde{\epsilon}/2\sqrt{2}$.  The formula for the deviation of
$\delta$ from $\pi/2$ is more complicated. In the case of normal
hierarchy, we obtain
$$
\cos \delta\sim \tilde{\epsilon} \left({\Delta m_{31}^2 \over
\Delta m_{21}^2}\right)^{1/2}\   \ll 1.
$$

\subsection{$A_4$ symmetry and maximal CP violation}

The  $\mu-\tau$ reflection symmetry can  be implicit or accidental
in models with certain flavor symmetries. We  show this using the
example of a specific model based on the $A_4$ symmetry. It has
been observed that in  models based on $A_4$, the Dirac phase is
maximal~\cite{A4first}. The neutrino mass matrix in the flavor
basis is given by \be {m}_\nu= f^2 \langle H\rangle^2 U_L^T
M_N^{-1} U_L, \label{a4matr} \ee where
\begin{eqnarray} U_L = \left[ \matrix{ 1 & 1 &1 \cr 1 &
e^{2i\pi/3} & e^{-2i\pi/3}\cr 1 & e^{-2i\pi/3} & e^{2i\pi/3}
}\right],
\end{eqnarray} and
 $f$ is the Yukawa coupling, and in the symmetry basis
the Dirac mass matrix is proportional to the unit matrix.

In the limit of exact $A_4$ one has $M_N= {\rm Diag}(M,M,M)$,  and
as is shown in \cite{A4first}, the corresponding light neutrino
mass matrix,
\begin{eqnarray}
{m}_\nu= \frac{f^2 \langle H\rangle^2}{M} \left[ \matrix{1 & 0 & 0
\cr 0 & 0 & 1 \cr 0& 1 & 0}\right]
\end{eqnarray}
leads to a degenerate mass spectrum. Deviations of  $M_N$ from
being proportional to the unit matrix, which can take place due to
the soft breaking of the $A_4$ symmetry or due to the radiative
corrections, can lead to phenomenologically acceptable mass matrix
for light neutrinos~\cite{A4first}.

Let us show that for {\it any real} (and non-singular) matrix
$M_N$, the Dirac phase produced by $m_\nu$ (\ref{a4matr}) is
maximal, $\delta = \pi/2$. Indeed,  in the flavor basis,  the
Dirac mass matrix of neutrinos is given by   $m_D \equiv f \langle
H\rangle U_L$.  It is easy to check that in this case both the
Dirac and the Majorana mass terms ($M_N$ is real)  are invariant
under the $\mu-\tau$ reflection symmetry (\ref{defsym}) with \be
N_i \to N_i^c. \ee The whole neutrino sector will therefore be
invariant under the $\mu-\tau$ reflection and this, in turn, leads
to the maximal Dirac phase.

\section{Nonzero CP violating phase \label{nonzerosection}}
In this section, we introduce a particular form of CP-flavor
transformation that can be considered as a generalization of the
$\mu-\tau$ reflection transformations. We then discuss how this
symmetry can be realized within the framework of the seesaw
mechanism and discuss its implications for the leptogenesis.

\subsection{Generalized $\mu -  \tau$ reflection symmetry\label{generalized}}

Let us introduce the following CP-flavor transformations: \be
\nu_\alpha \to \sum_\beta P_{\alpha \beta}\nu_\beta^C,
\label{general} \ee where $P$ is a unitary matrix and $\alpha$ and
$\beta$ are flavor indices. We consider consequences  of the
invariance of the neutrino mass terms with respect to these
transformations. Apparently (\ref{general}) is a generalization of
the $\mu - \tau$ reflection.

Below we  define $P$ in a specific way that implies a definite
value for the Dirac phase which depends on parameters of
transformation (\ref{general}). Let us consider the following
symmetric form for $P$ \be
{P}(\theta_{23},\phi)=U_{23}(\theta_{23}){\rm
Diag}[1,1,e^{i\phi}]U_{23}^T(\theta_{23}). \label{specific} \ee
Notice that in the  case $e^{i\phi}=-1$,  the matrix ${P}$ equals
\ba {P}=\left[\matrix{1&0&0\cr 0 & 0&-1 \cr 0&-1&0}\right] \ea
which corresponds to the $\mu\ - \ \tau$ reflection with
$\xi_1=-\xi_2=-\xi_3=1$ [see Eq.~(\ref{defsym})].

Since $U_{12}$ and ${\rm Diag}[1,1,e^{i\phi}]$ commute,  it is
straightforward to show that
$$
{P(\theta_{23}, \phi)}\cdot
U_{PMNS}(\theta_{23},\theta_{12},\theta_{13},\delta,\phi_2,\phi_3)=
U_{PMNS}(\theta_{23},\theta_{12},\theta_{13},\delta+\phi,\phi_2,\phi_3+2\phi).
$$
Thus, the equality \be {P}(\tilde{\theta}_{23},\phi) \cdot m_\nu
\cdot {P}(\tilde{\theta}_{23},\phi)=m_\nu^* \label{equalP} \ee is
satisfied if \be \theta_{23}=\tilde{\theta}_{23} \ee and \be
\delta= - \frac{\phi}{2}, \ \ \ \phi_2=0~ {\rm or}~ \pi, \ \ \
\phi_3=-\phi. \label{phaseequal}
\ee
Eq.~(\ref{equalP}) implies that the mass matrix $m_\nu$ is
invariant under the CP-flavor transformation (\ref{general}) with
$P(\theta_{23}, \phi)$ given in (\ref{specific}) only if the
CP-violating phases associated with $m_\nu$ satisfy equalities
(\ref{phaseequal}).  From (\ref{phaseequal}) we obtain a simple
relation between the Dirac and Majorana phases: \be \phi_3 =
2\delta. \label{twice}\ee
Notice that, here, $\sin \theta_{13}$  can  take any real value
(including negative values) but $\delta$ is restricted to
$[0,\pi]$, so the prediction is free from any ambiguity.

Let us consider some special cases:

1)  To achieve a maximal Dirac phase and a small deviation of
$\theta_{23}$ from its maximal mixing value, the transformation
matrix should  be \ba
\hat{P}[\delta=\pi/2,\theta_{23}=\pi/4+\tilde{\theta}]=\left[
\matrix{1&0&0 \cr 0 & -2\tilde{\theta} & -1\cr
0&-1&2\tilde{\theta}}\right],\ea where $\tilde{\theta}\ll 1$.

2) Maximal mixing, $\theta_{23}=\pi/4$, and  a given value of the
Dirac phase $\delta$, can be achieved with \ba {P}=\left[\matrix{1
& 0& 0\cr 0 &{e^{-2i\delta}+1 \over \sqrt{2}}& {e^{-2i\delta}-1
\over \sqrt{2}}\cr 0& {e^{-2i\delta}-1 \over
\sqrt{2}}&{e^{-2i\delta}+1 \over \sqrt{2}}}\right]. \ea

In summary, the symmetry under the CP-flavor transformation
defined in (\ref{specific}) leads to certain values of the Dirac
and Majorana CP-violation phases (in the standard
parametrization). The phases are related to the parameter of
transformation and, depending on the value of this parameter, can
take any value from zero to the maximal CP-violating phase.

\subsection{Seesaw mechanism and leptogenesis \label{leptogenesis}}

Let us consider  the implications of the neutrino mass matrices
with the CP-flavor symmetry for the seesaw mechanism and
leptogenesis. We introduce three  heavy right-handed neutrinos
$N_i$, and consider a basis in which the mass matrix of these
neutrinos is diagonal: $M = Diag(M_1, M_2, M_3)$.
As discussed in sec. (\ref{cpreflection}), if the right-handed
neutrinos transform under the $\mu-\tau$ reflection without any
flavor change, $N_i \to N_i^c$, $M_i$ should be real and the
Yukawa coupling matrix should have  form (\ref{yukkk}) with real
$a_i$ and $b_i$. For $Y_{\nu}$ given in Eq.~(\ref{yukkk}) $Y_\nu
\cdot Y_\nu^\dagger $ is a real matrix. This conclusion is valid
also for the generalized $\mu -\tau$ reflection with a more
complicated transformation matrix ${P}$: Invariance under $N_i \to
N_i^c$ and $\nu \to {P}\nu^c$ implies $Y_\nu^*=Y_\nu\cdot
\hat{P}$. As a result,
$$Y_\nu \cdot Y_\nu^\dagger=Y_\nu \hat{P} \cdot \hat{P}^\dagger
Y_\nu^\dagger= Y_\nu^*\cdot Y_\nu^T.$$ Leptogenesis is driven by a
combination of the Yukawa couplings known as the total asymmetry
($\varepsilon_1$)  which is given by ${\rm Im}[(Y_\nu\cdot
Y_\nu^\dagger)_{\beta 1}^2]$. Obviously, the total asymmetry
vanishes for a model which is  symmetric under the $\mu-\tau$
reflection and, without  flavor effects associated with the
charged leptons, leptogenesis cannot take place in the symmetry
limit. This result is in accord with \cite{grimus}.

However, as recently shown, for $M_1<10^{13}$~GeV  flavor effects
can alter the situation; {\it i.e.,} as long as the partial
asymmetry \be \varepsilon_1^\alpha\equiv \left[\Gamma(N_1\to
\ell_\alpha H)-\Gamma(N_1\to
\bar{\ell}_\alpha\bar{H})\right]/\Gamma_{{\rm tot}}
\label{partial-asymmetry} \ee does not vanish, decoherence caused
by charged lepton Yukawa couplings of $\tau$ can lead to a
successful leptogenesis even if $\varepsilon_1=0$ \cite{nir}. In
general, the mass matrix in Eq.~(\ref{yukkk}) yields nonzero
$\varepsilon_1^\tau$ which for $M_1<10^{13}$~GeV can reopen the
possibility of a successful leptogenesis.

Another possibility is to  define the transformation of the $N_i$
in a way that includes  a flavor permutation: \be N_1
\leftrightarrow N_2^c, ~~~ N_3 \to N_3^c. \label{ncp12} \ee
Invariance under these transformations imply \ba Y_\nu =\left[
\matrix{a_1 & b_1 & b_2^* \cr a_1^* & b_2 & b_1^*
 \cr a_3 & b_3 &b_3^* }\right],
\ea where $a_3$ is real but the rest of the parameters can be
complex. Now the  matrix   $Y_\nu\cdot Y_\nu^\dagger$ has complex
off-diagonal entries which opens a possibility for leptogenesis.
Moreover the transformations defined in (\ref{ncp12}) imply
$$M_1=M_2.$$
This degeneracy can be slightly lifted by some additional physics
\cite{ahn} (notice that the $\mu-\tau$ reflection symmetry is
broken anyway in the charged lepton sector). The quasi-degeneracy
of these two mass eigenstates can lead to the resonance
leptogenesis.

\section{Possible relations between the  phases of the CKM and  PMNS matrices}

In view of strong differences between the  mass and mixing
patterns  of  leptons and quarks, one  does not expect
$\delta_{CKM}$ and $ \delta$ to be equal. However, conditions can
be formulated that  lead to simple and immediate relations of the
phases. For this purpose, we first make the following assumptions:

1) The seesaw type-I mechanism generates the  neutrino masses,
and therefore  the light neutrino mass matrix in the flavor basis
is equal to \be m_{\nu} = U_L^* m_D^{diag} M^{-1}_N m_D^{diag}
U_L^{\dagger}; \label{seesawm} \ee

2)  Due to the quark-lepton symmetry or unification \be U_{L} =
V_{CKM}^\dagger; \ee

3)  The matrix $m_D^{diag} M^{-1}_N m_D^{diag}$ in (\ref{seesawm})
is diagonalized by a bi-maximal rotation \cite{bm} \be U_{bm} =
U_{23}^m U_{12}^m, \ee where $U_{ij}^m$ is the maximal ($\pi/4$)
rotation in the $ij-$ plane.


Then, the lepton mixing matrix will be equal to \be U_{PMNS} =
V_{CKM}^{\dagger} U_{bm}, \label{matrrel} \ee
which  leads to acceptable  values for mixing angles according to
the quark-lepton complementarity (QLC) scenario \cite{qlc}. From
(\ref{matrrel}) we obtain   \be \label{13} |\sin\theta_{13}| =
\frac{1}{\sqrt{2}} \left| V_{td}^{\dagger} +
V_{cd}^{\dagger}\right| \simeq \frac{\sin \theta_C}{\sqrt{2}}.\ee
This relation can be tested by  forthcoming experiments such as
Double CHOOZ  \cite{chooz}, Daya Bay \cite{DayaBay}, T2K
\cite{T2K} and NO$\nu$A \cite{nova}.
Calculating the Jarlskog invariant and inserting the values of the
mixing angles, we  find \be \sin \delta \approx
\frac{|V_{ub}|}{\sin\theta_C} \sin \delta_{CKM}.
\label{leptoquark1} \ee 
Here $V_{ub}$ is an element of the CKM matrix and $\theta_C$ is
the Cabibbo angle. This leads to a suppressed value for the Dirac
phase.
Inserting the best fit values of  $|V_{ub}| $ and $\sin \theta_C$
\cite{pdg} in (\ref{leptoquark1}), we find
$\delta=(0.97^{+0.10}_{-0.12})^\circ$ where the uncertainty results
from the relatively large uncertainty in $\delta_{CKM}$.

The Lagrangian of the quark sector is invariant under $V_{CKM} \to
\Gamma_\phi^\dagger V_{CKM},$ where $\Gamma_\phi$ is a diagonal
matrix whose eigenvalues are pure phases. However, $U_{PMNS}$
given by (\ref{matrrel}) changes non-trivially under this
transformation. This results in an ambiguity in evaluation of
$\delta$.
 In general,  the seesaw
mechanism can lead to the appearance of an additional phase matrix
\be U_{PMNS} = V_{CKM}^{\dagger} \Gamma_{\phi} U_{bm}.
\label{matrrel3} \ee  Including the phase matrix $\Gamma_{\phi}$,
the leptonic phase can be much larger than (\ref{leptoquark1});
however, the value of the phase should be restricted in order to
make (\ref{matrrel3}) compatible with the
data on the mixing angles \cite{antush-king}. \\

Let us consider another possibility that also agrees with the
data. The bimaximal mixing can be  generated by the charged lepton
mass matrix. In this case \be U_{PMNS} =  U_{bm}
V_{CKM}^{\dagger}, \label{matrrel2} \ee which leads to \be
\sin\theta_{13}  = \frac{|V_{td}^{\dagger}  +
V_{ts}^{\dagger}|}{\sqrt{2}} \simeq \frac{|V_{cb}|}{\sqrt{2}}\ee
and \be \sin \delta \approx -\frac{|V_{ub}|}{V_{cb}} \sin
\delta_{CKM} . \ee Unfortunately, such a small value of $\sin
\theta_{13}$ will be beyond the reach of the forthcoming
experiments designed to measure $\sin \theta_{13}$
\cite{chooz,DayaBay,T2K}. Thus, a positive result in these
experiments will exclude this possibility.

Like in the case of (\ref{matrrel}), the lepton phase is
suppressed: $\delta\simeq  5^{\circ}$. Thus, we conclude that
without introducing new phases, the  immediate relations between
the quark and lepton phases lead to suppression of the leptonic
CP-phase in comparison with the quark phase.
Essentially, this is a consequence of the large lepton mixing.\\


\section{Conclusion}

We have   studied symmetries, principles and phenomenological
conditions which entail certain values for the Dirac CP-violating
phase in the leptonic sector. Such a study gives some idea about
physics behind the CP-violation as well as the implications of
future measurements of the phase.

Bearing in mind that even in the quark sector, there is no
theory of CP-violation, we have considered the following
possibilities:

- zero (or a very small) phase;

- a maximal CP-violating phase, $\delta = \pi/2$;

- an arbitrary phase which depends on the parameter of symmetry
transformation;

- certain relation between the phases in  the quark and lepton
sectors.\\

By defining rephasing invariant
combinations of the elements of the neutrino mass matrix, we have
formulated the necessary and sufficient conditions for the zero
value of the phase.
In the case that all the
elements of $m_\nu$ are nonzero, CP-invariance of the mass matrix
is equivalent to the realness of the three rephashing invariant
combinations $I_1$, $I_2$ and $I_5$ defined in
(\ref{rephasinginvariant},\ref{I5}). In other words, if  $I_1$,
$I_2$ and $I_5$ are all real, the Dirac as well as Majorana
CP-violating phases will be zero (or equal to $\pi$).
Particular cases in which  some of the elements of $m_\nu$ are zero
have also been discussed. \\

 We have  studied the possibility that the Dirac
phase is zero or very small but the Majorana phases are sizeable;
{\it i.e.,}  CP is still broken in the lepton sector of the theory
despite the vanishing Jarlskog invariant. We have derived the
general form of the mass matrix that satisfies these conditions
[see (\ref{real-off-diag})]. There is no unique symmetry which
leads to such a form; however, we have found that by changing the
basis, the matrix can be written in a simple form [see
(\ref{newbasis},\ref{simpleinnew})] which will be easier to
incorporate in models. We have observed that the symmetries and
mass patterns that lead to zero $\delta$
  also yield certain relations between
the 1-3 mixing and  other observables [see
Eqs.~(\ref{13pred},\ref{predi},\ref{another13})].
These relations can be used as a test  for the underlying
physics. \\

 Maximal Dirac CP-violating   phase can be related to a symmetry
under a specific type of combined  CP and flavor (CP-flavor)
transformations
that is known as the $\mu-\tau$ reflection symmetry.
 The symmetry
leads to $\delta = \pi/2$, zero (or  equal to $\pi$) Majorana
phases and maximal $\nu_{\mu} - \nu_{\tau}$ mixing.
We have shown that this symmetry is a sufficient (if mass matrix is complex) but not
a necessary condition for $\delta = \pi/2$. In order to verify this symmetry,
in addition to confirming $\cos \delta=\cos 2 \theta_{23}=0$, it is necessary to
check that
the Majorana phases are zero (or equal to $\pi$).\\

We have proposed a generalized version of the $\mu-\tau$
reflection symmetry.   Depending on the value of the parameter of
transformation, this symmetry can lead to any value of the Dirac
phase. This symmetry predicts a simple relation between the
Majorana and Dirac phases [see (\ref{phaseequal},\ref{twice})]. A
mass matrix symmetric under the CP-flavor transformation can be
generated within the seesaw mechanism. We have discussed
leptogenesis in the context of
 seesaw mechanism respecting the generalized $\mu-\tau$ symmetry.
If  the right-handed neutrinos transform into CP-conjugate of
themselves under this symmetry, $N_i \to N_i^C$, the total asymmetry
vanishes; however, if the flavor effects are taken into account,
the successful leptogenesis can still be realized. The successful
leptogenesis can also be obtained in the case of non-trivial
(flavor) transformation for $N_i$. In this case a weak violation
of the symmetry can lead to the resonant leptogenesis.

 The leptonic phase can be related to the quark phase in the
context of quark-lepton complementarity. In this scenario, the
mixing matrix in the lepton sector appears as a combination of the
CKM mixing and a bi-maximal mixing. If no additional phase  apart
from the CKM phase is introduced, one expects a suppressed Dirac
phase as a consequence of the large lepton mixing.

\section*{Appendix}

The matrix $h$ (\ref{hmatr}) can be written as

\be h= m_1^2{\rm Diag}[1, \ 1, \ 1]+ U_{{\rm PMNS}}\cdot {\rm
Diag}[0,\ \Delta m_{21}^2, \ \Delta m_{31}^2] \cdot U_{\rm
PMNS}^\dagger, \ee where $\Delta m_{21}^2 \equiv m_2^2-m_1^2$ and
$\Delta m_{31}^2\equiv m_3^2-m_1^2$ and $U_{PMNS}$ is the mixing
matrix in the standard  parametrization defined in \cite{pdg}. It
is straightforward to show that
$$
h_{ee}=m_1^2+\Delta m_{21}^2s_{12}^2c_{13}^2+\Delta
m_{31}^2s_{13}^2,
$$
$$
h_{\mu \mu}=m_1^2+\Delta m_{31}^2s_{23}^2c_{13}^2+ \Delta
m_{21}^2\left(c_{12}^2c_{23}^2+s_{12}^2s_{13}^2s_{23}^2
-2s_{13}s_{12}c_{12}s_{23}c_{23}\cos \delta\right),
$$
$$
h_{\tau \tau}=m_1^2 +\Delta m_{31}^2c_{23}^2c_{13}^2+\Delta
m_{21}^2 \left(
s_{23}^2c_{12}^2+s_{12}^2c_{23}^2s_{13}^2+2s_{12}c_{12}s_{23}c_{23}s_{13}\cos
\delta \right).
$$
The absolute values of the off-diagonal elements are as follows
  \be |h_{\mu \tau}| =\left| s_{23}c_{23} \left[
  \Delta m_{31}^2c_{13}^2+
\Delta m_{21}^2\left[(s_{12}^2s_{13}^2-c_{12}^2)+
s_{12}c_{12}s_{13}(s_{23}^2e^{i
\delta}-c_{23}^2e^{-i\delta})\right]\right]\right|,
\label{hmutau}\ee \be |h_{e\mu}|= \left|\Delta
m_{31}^2s_{23}s_{13}c_{13}e^{-i
  \delta}+\Delta m_{21}^2s_{12}c_{13}\left(c_{12}c_{23}-s_{12}s_{13}s_{23}e^{-i
  \delta}\right)\right|, \label{hemu}
\ee \be |h_{e\tau}| =\left| \Delta m_{31}^2c_{23}s_{13}c_{13}e^{-i
  \delta}+\Delta m_{21}^2s_{12}c_{13}\left(-c_{12}s_{23}-s_{12}s_{13}c_{23}e^{-i
  \delta}\right)\right|.
\label{hetau} \ee

\section*{Acknowledgments}

Y. F. is grateful to   the Abdus Salam International Centre for
Theoretical Physics (ICTP) where  this work started, for generous
hospitality.


\begin{thebibliography}{99}

\bibitem{cp}
J.~H.~Christenson, J.~W.~Cronin, V.~L.~Fitch and R.~Turlay,
  Phys.\ Rev.\ Lett.\  {\bf 13}, 138 (1964);
  B.~Aubert {\it et al.}  [BABAR Collaboration],
  Phys.\ Rev.\ Lett.\  {\bf 87}, 091801 (2001)
  [arXiv:hep-ex/0107013].
\bibitem{factory} A.~De Rujula, M.~B.~Gavela and P.~Hernandez,
  Nucl.\ Phys.\ B {\bf 547} (1999) 21
  [arXiv:hep-ph/9811390]; A.~Cervera, A.~Donini, M.~B.~Gavela, J.~J.~Gomez Cadenas, P.~Hernandez, O.~Mena and S.~Rigolin,
  Nucl.\ Phys.\ B {\bf 579} (2000) 17
  [Erratum-ibid.\ B {\bf 593} (2001) 731]
  [arXiv:hep-ph/0002108];
   C.~Albright {\it et al.},
  arXiv:hep-ex/0008064;
  K.~Dick, M.~Freund, M.~Lindner and A.~Romanino,
  Nucl.\ Phys.\ B {\bf 562}, 29 (1999)
  [arXiv:hep-ph/9903308];
  V.~D.~Barger, S.~Geer, R.~Raja and K.~Whisnant,
  Phys.\ Rev.\ D {\bf 63}, 033002 (2001)
  [arXiv:hep-ph/0007181];
 M.~Freund, P.~Huber and M.~Lindner,
  Nucl.\ Phys.\ B {\bf 585}, 105 (2000)
  [arXiv:hep-ph/0004085];
  K.~Hagiwara, N.~Okamura and K.~i.~Senda,
  arXiv:hep-ph/0607255.

\bibitem{A4CP}
 K.~S.~Babu, E.~Ma and J.~W.~F.~Valle,
  Phys.\ Lett.\ B {\bf 552}, 207 (2003)
  [arXiv:hep-ph/0206292].

\bibitem{yasue}
I.~Aizawa and M.~Yasue,
  Phys.\ Lett.\ B {\bf 607}, 267 (2005)
  [arXiv:hep-ph/0409331];
  I.~Aizawa, T.~Kitabayashi and M.~Yasue,
  Phys.\ Rev.\ D {\bf 72}, 055014 (2005)
  [arXiv:hep-ph/0504172].
  I.~Aizawa, T.~Kitabayashi and M.~Yasue,
  %
  Nucl.\ Phys.\ B {\bf 728}, 220 (2005)
  [arXiv:hep-ph/0507332];
I.~Aizawa and M.~Yasue,
  %
  Phys.\ Rev.\ D {\bf 73}, 015002 (2006)
  [arXiv:hep-ph/0510132].
\bibitem{texturezero}
 K.~Matsuda and H.~Nishiura,
  Phys.\ Rev.\ D {\bf 74}, 033014 (2006)
  [arXiv:hep-ph/0606142];
S.~Kaneko, H.~Sawanaka and M.~Tanimoto,
  JHEP {\bf 0508}, 073 (2005)
  [arXiv:hep-ph/0504074];
  Z.~z.~Xing,
  Phys.\ Lett.\ B {\bf 530}, 159 (2002)
  [arXiv:hep-ph/0201151];

\bibitem{Jarlskog}
C.~Jarlskog,
  %
  Phys.\ Rev.\ Lett.\  {\bf 55}, 1039 (1985).
\bibitem{shrock}
  A.~Kusenko and R.~Shrock,
  arXiv:hep-ph/9403315.
\bibitem{recenthindu}
    U.~Sarkar and S.~K.~Singh,
  arXiv:hep-ph/0608030.

\bibitem{yasaman}Y. Farzan, Talk given at the XXXIII International Conference
on High Energy Physics, July 26 - August 2, 2006 Moscow, Russia,
http://ichep06.jinr.ru/reports/92$\_$2s3$\_$15p15$\_$farzan.pdf.
\bibitem{chooz}
  P.~Huber, J.~Kopp, M.~Lindner, M.~Rolinec and W.~Winter,
  JHEP {\bf 0605} (2006) 072
  [arXiv:hep-ph/0601266].
  \bibitem{DayaBay}
  J.~Cao,
  Nucl.\ Phys.\ Proc.\ Suppl.\  {\bf 155} (2006) 229
  [arXiv:hep-ex/0509041].
  \bibitem{T2K}
  Y.~Itow {\it et al.},
  arXiv:hep-ex/0106019.

\bibitem{nova}
D.~S.~Ayres {\it et al.}  [NOvA Collaboration],
  arXiv:hep-ex/0503053.

\bibitem{pdg}
W.-M. Yao et al., Particle Data Group, J. Phys. G 33, 1 (2006).

\bibitem{mohapatra}
 R.~N.~Mohapatra and W.~Rodejohann,
  %
  Phys.\ Rev.\ D {\bf 72}, 053001 (2005)
  [arXiv:hep-ph/0507312].
\bibitem{xing}
  Z.~z.~Xing, H.~Zhang and S.~Zhou,
  arXiv:hep-ph/0607091.
\bibitem{triangle}
 Y.~Farzan and A.~Y.~Smirnov,
  Phys.\ Rev.\ D {\bf 65}, 113001 (2002)
  [arXiv:hep-ph/0201105];
J.~F.~Beacom, N.~F.~Bell, D.~Hooper, S.~Pakvasa and T.~J.~Weiler,
  Phys.\ Rev.\ D {\bf 69}, 017303 (2004)
  [arXiv:hep-ph/0309267].
   W.~Winter,
  Phys.\ Rev.\ D {\bf 74}, 033015 (2006)
  [arXiv:hep-ph/0604191].

\bibitem{flavorratio}
 J.~F.~Beacom, N.~F.~Bell, D.~Hooper, S.~Pakvasa and T.~J.~Weiler,
  Phys.\ Rev.\ D {\bf 68}, 093005 (2003)
  [Erratum-ibid.\ D {\bf 72}, 019901 (2005)]
  [arXiv:hep-ph/0307025].

\bibitem{scott}
P.~F.~Harrison and W.~G.~Scott,
  Phys.\ Lett.\ B {\bf 547}, 219 (2002)
  [arXiv:hep-ph/0210197].



\bibitem{grimus} W.~Grimus and L.~Lavoura,
  Phys.\ Lett.\ B {\bf 579}, 113 (2004)
  [arXiv:hep-ph/0305309].
  \bibitem{takeshi}
  T.~Fukuyama and H.~Nishiura,
  arXiv:hep-ph/9702253.
  \bibitem{king}
    S.~F.~King,
  Phys.\ Lett.\ B {\bf 439}, 350 (1998)
  [arXiv:hep-ph/9806440].



\bibitem{A4first}

E.~Ma and G.~Rajasekaran,
  Phys.\ Rev.\ D {\bf 64}, 113012 (2001)
  [arXiv:hep-ph/0106291].


\bibitem{nir}
 A.~Abada, S.~Davidson, F.~X.~Josse-Michaux, M.~Losada and A.~Riotto,
  JCAP {\bf 0604}, 004 (2006)
  [arXiv:hep-ph/0601083];
 A.~Abada, S.~Davidson, A.~Ibarra, F.~X.~Josse-Michaux, M.~Losada and A.~Riotto,
  JHEP {\bf 0609}, 010 (2006)
  [arXiv:hep-ph/0605281];
 E.~Nardi, Y.~Nir, E.~Roulet and J.~Racker,
  JHEP {\bf 0601}, 164 (2006)
  [arXiv:hep-ph/0601084];
{\it see also,} R.~Barbieri {\it et al.},
  Nucl.\ Phys.\ B {\bf 575}, 61 (2000)
  [arXiv:hep-ph/9911315];
  T.~Endoh, T.~Morozumi and Z.~h.~Xiong,
  Prog.\ Theor.\ Phys.\  {\bf 111}, 123 (2004)
  [arXiv:hep-ph/0308276];
  T.~Fujihara {\it et al.,}
  Phys.\ Rev.\ D {\bf 72}, 016006 (2005)
  [arXiv:hep-ph/0505076];
  S.~Antusch, S.~F.~King and A.~Riotto,
  arXiv:hep-ph/0609038;
   O.~Vives,
  Phys.\ Rev.\ D {\bf 73}, 073006 (2006)
  [arXiv:hep-ph/0512160];
S.~Pascoli, S.~T.~Petcov and A.~Riotto,
  arXiv:hep-ph/0609125;
 G.~C.~Branco, R.~Gonzalez Felipe and F.~R.~Joaquim,
  arXiv:hep-ph/0609297.

\bibitem{ahn}
K.~Turzynski,
  Phys.\ Lett.\ B {\bf 589}, 135 (2004)
  [arXiv:hep-ph/0401219];
Y.~H.~Ahn, C.~S.~Kim, S.~K.~Kang and J.~Lee,
  arXiv:hep-ph/0610007;
  R.~Gonzalez Felipe, F.~R.~Joaquim and B.~M.~Nobre,
  Phys.\ Rev.\ D {\bf 70}, 085009 (2004)
  [arXiv:hep-ph/0311029];
 G.~C.~Branco, R.~Gonzalez Felipe, F.~R.~Joaquim and B.~M.~Nobre,
  Phys.\ Lett.\ B {\bf 633}, 336 (2006)
  [arXiv:hep-ph/0507092].



\bibitem{bm}
    V.~D.~Barger, S.~Pakvasa, T.~J.~Weiler and K.~Whisnant,
  Phys.\ Lett.\ B {\bf 437}, 107 (1998)
  [arXiv:hep-ph/9806387].



\bibitem{qlc}
  M.~Raidal,
  Phys.\ Rev.\ Lett.\  {\bf 93}, 161801 (2004)
  [arXiv:hep-ph/0404046];
  H.~Minakata and A.~Y.~Smirnov,
  Phys.\ Rev.\ D {\bf 70}, 073009 (2004)
  [arXiv:hep-ph/0405088];
  J.~Ferrandis and S.~Pakvasa,
  Phys.\ Rev.\ D {\bf 71}, 033004 (2005)
  [arXiv:hep-ph/0412038];
S.~K.~Kang, C.~S.~Kim and J.~Lee,
  Phys.\ Lett.\ B {\bf 619}, 129 (2005)
  [arXiv:hep-ph/0501029];
  N.~Li and B.~Q.~Ma,
  Phys.\ Rev.\ D {\bf 71}, 097301 (2005)
  [arXiv:hep-ph/0501226];
K.~Cheung, S.~K.~Kang, C.~S.~Kim and J.~Lee,
  Phys.\ Rev.\ D {\bf 72}, 036003 (2005)
  [arXiv:hep-ph/0503122];
  Z.~z.~Xing,
  Phys.\ Lett.\ B {\bf 618}, 141 (2005)
  [arXiv:hep-ph/0503200];
  A.~Datta, L.~Everett and P.~Ramond,
  Phys.\ Lett.\ B {\bf 620}, 42 (2005)
  [arXiv:hep-ph/0503222].




\bibitem{antush-king}
  S.~Antusch and S.~F.~King,
  Phys.\ Lett.\ B {\bf 631}, 42 (2005)
  [arXiv:hep-ph/0508044].


\end{thebibliography}
\end{document}